\documentclass[final,3p,times]{elsarticle}

\usepackage{wrapfig}

\usepackage{graphics}
\usepackage{graphicx}
\usepackage{epsfig}
\usepackage{amssymb}
\usepackage{amsthm}
\usepackage{amsmath}

\usepackage{caption}
\usepackage[labelfont=bf]{caption}
\usepackage[labelsep=space]{caption}
\usepackage[nodots, compress]{numcompress}
\biboptions{sort&compress}

\usepackage{xcolor}
\usepackage{hyperref}
\hypersetup{
colorlinks,
citecolor=[violet],
linkcolor=[red],
urlcolor=[blue]
}

\journal{Nano Energy}

\begin{document}

\begin{frontmatter}

\title{Integrating Machine Learning with Triboelectric Nanogenerators: Optimizing Electrode Materials and Doping Strategies for Intelligent Energy Harvesting}

\author[1]{Guanping Xu\fnref{fn1}}
\author[1]{Zirui Zhao\fnref{fn1}}
\fntext[fn1]{These authors contributed equally.}
\author[2]{Zhong Lin Wang}
\author[1]{Hai-Feng Li\corref{cor1}}
\cortext[cor1]{Corresponding author}
\ead{haifengli@um.edu.mo}

\affiliation[1]{organization={Institute of Applied Physics and Materials Engineering, University of Macau},
            addressline={Avenida da Universidade, Taipa},
            city={Macao SAR},
            postcode={999078},
            country={China}}

\affiliation[2]{organization={Beijing Key Laboratory of Micro-Nano Energy and Sensor, Center for High-Entropy Energy and Systems, Beijing Institute of Nanoenergy and Nanosystems, Chinese Academy of Sciences},
            addressline={},
            city={Beijing},
            postcode={101400},
            country={China}}

\clearpage
      
\begin{abstract}
The integration of machine learning techniques with triboelectric nanogenerators (TENGs) offers a transformative pathway for optimizing energy harvesting technologies. In this study, we propose a comprehensive framework that utilizes graph neural networks to predict and enhance the performance of TENG electrode materials and doping strategies. By leveraging an extensive dataset of experimental and computational results, the model effectively classifies electrode materials, predicts optimal doping ratios, and establishes robust structure-property relationships. Key findings include a 65.7\% increase in energy density for aluminum-doped PTFE and an 85.7\% improvement for fluorine-doped PTFE, highlighting the critical influence of doping materials and their concentrations. The model further identifies PTFE as a highly effective negative electrode material, achieving a maximum energy density of 1.12~J/cm$^2$ with 7\% silver (Ag) doping when copper (Cu) is used as the positive electrode. This data-driven approach not only accelerates material discovery but also significantly reduces experimental costs, providing novel insights into the fundamental factors influencing TENG performance. The proposed methodology establishes a robust platform for intelligent material design, advancing the development of sustainable energy technologies and self-powered systems.
\end{abstract}

\begin{graphicalabstract}
\includegraphics[width=0.88\textwidth]{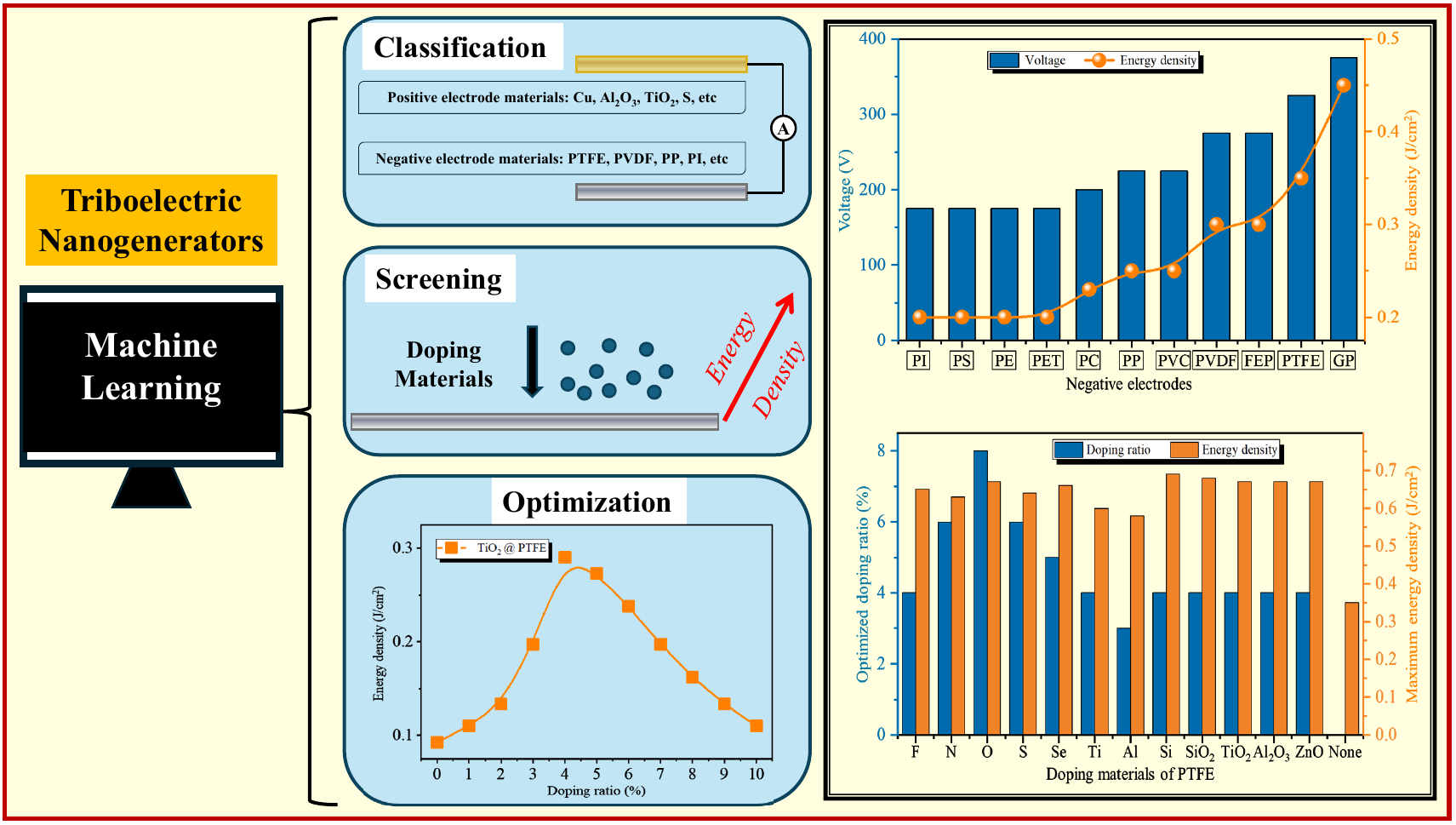} \newline
\noindent{\textbf{Caption of Graphical Abstract:} This study integrates machine learning techniques, including graph neural networks, to optimize electrode materials and doping strategies for triboelectric nanogenerators. By leveraging a data-driven model, it predicts material performance with high accuracy, enhances energy density, and accelerates material discovery. The findings establish a robust foundation for intelligent energy harvesting and sustainable energy technologies.}
\end{graphicalabstract}



\end{frontmatter}

\section{Introduction}

With global industrialization and the increasing demand for energy, the need for efficient energy harvesting technologies has grown substantially \cite{mcelroy2009potential, lewis2007toward, moral2021measuring, hasan2023harnessing}. Among these technologies, the triboelectric nanogenerator (TENG) has garnered significant attention due to its low cost, lightweight design, flexibility, and versatility in material selection \cite{fan2012flexible, wang2017triboelectric, fan2016flexible, wang2012nanotechnology}. The performance and power output of TENGs are largely determined by the selection of positive and negative electrode materials, as well as their surface properties \cite{YANG201941, WANG201934, xu2019contact}. Additionally, variations in the types and proportions of dopants in electrodes significantly affect the overall performance of TENGs \cite{zhu2013toward, ARKAN2023102683}. As the demand for intelligent self-powered systems based on TENG technology continues to rise, there is a pressing need to enhance output performance \cite{cima2014next, wang2016sustainably}. To address these challenges, the integration of machine learning (ML) techniques offers a promising solution for optimizing TENG material selection.

ML, first introduced in 1959, has experienced remarkable growth in recent years \cite{carbonell1983machine}, driven by advancements in computational power. Experts across diverse fields—including game development, economics, and bioinformatics—have successfully integrated ML, achieving impressive outcomes. With the widespread availability of ML resources and tools, coupled with enhanced data processing capabilities, ML holds significant potential to reduce material screening time and costs \cite{moosavi2020role, fare2022multi, cai2020machine}. Unlike traditional methods that rely heavily on manual material selection and performance testing, the characterization and performance analysis of TENGs often require substantial time and human resources for design and experimental validation. This growing demand for efficient material optimization has driven increased interest in ML applications for TENG research. While ML has already demonstrated notable progress in integrating nanotechnology with energy systems \cite{yang2024triboelectric, lu2022decoding, liu2024self, jin2020triboelectric}, a comprehensive predictive model to guide the selection of optimal positive and negative electrode materials for TENGs remains underdeveloped. 

Graph Neural Networks (GNNs) are particularly suited for TENG material optimization due to their ability to model complex, non-linear relationships between material components, such as electronegativity and dielectric constants. Traditional ML methods, including fully connected neural networks and support vector machines, often struggle to capture these intricate dependencies. In contrast, GNNs incorporate structural and relational data, making them ideal for understading the interconnected properties of TENG materials.

Furthermore, unsupervised learning in GNNs enables the model to explore datasets without relying on limited labeled data, facilitating the discovery of new materials and doping strategies. This approach provides a scalable and efficient framework for material optimization, capable of uncovering hidden patterns and significantly accelerating the screening process \cite{yin2024ai}.

In this study, we present an innovative approach employing GNNs for predictive analysis and performance evaluation of TENG electrode materials. Our methodology establishes a comprehensive framework that systematically addresses the critical challenge of material optimization in TENG systems. The research encompasses three key phases: material classification, data acquisition and preprocessing, and predictive modeling. Initially, we implement a rigorous classification scheme for substrate materials and dopants associated with both positive and negative electrodes. This is followed by the compilation of an extensive dataset incorporating experimental measurements and computational simulations, capturing essential parameters such as parent material characteristics, dopant specifications, concentration ratios, and corresponding performance metrics \cite{zhu2013toward, ARKAN2023102683}. Through advanced data preprocessing techniques, we extract and quantify the dominant factors governing TENG output performance.

Traditional methods, such as experimental screening, first-principles calculations, and supervised ML, are often limited by their time-consuming nature, high computational costs, and reliance on labeled data. In contrast, our proposed approach, which combines GNN plus unsupervised learning, offers several key advantages. It effectively models complex material relationships and eliminates the dependency on labeled data, thereby enabling the exploration of a significantly broader material space. Despite its flexibility and efficiency, the method requires high-quality training data and faces challenges in interpretability when compared to traditional methods. Nonetheless, we believe this approach provides an innovative and scalable solution for optimizing TENG materials, demonstrating clear advantages over existing methods \cite{chen2024gnn, tosi202415}.

The core of our methodology lies in the development of a sophisticated GNN-based predictive model, which enables high-throughput screening of material combinations and precise performance prediction. This model eliminates the need for computationally intensive recalculations by establishing robust structure-property relationships. Utilizing unsupervised learning paradigms, the GNN architecture demonstrates exceptional capability in autonomous material classification. For optimal electrode material selection, our model validation reveals remarkable accuracy in predicting critical performance indicators, including output voltage, current density, and power generation, with particular emphasis on the influence of dopant characteristics and concentration profiles.

The systematic analysis conducted using our GNN model provides unprecedented insights into the key determinants of TENG performance, introducing a transformative approach to rational material selection and optimization for energy harvesting applications. This study marks a significant advancement in the field of intelligent material design for TENG systems by offering a robust computational platform that effectively links material properties with device performance.

By optimizing material selection, our approach enhances the efficiency and performance of TENGs, thereby improving energy harvesting in real-world applications. This underscores the practical value of our research in advancing energy solutions for wearable electronics and smart sensors \cite{wang2025omnidirectional}. Furthermore, the proposed methodology not only deepens our understanding of structure-performance relationships in TENGs but also establishes a new paradigm for data-driven material discovery in energy conversion technologies.

\section{Results and Discussion}

\subsection{Material classification model for TENGs}

A robust material classification model is essential for TENGs due to their exceptional versatility, environmentally sustainable characteristics, and potential for green energy applications. This framework must effectively classify electrode materials, evaluate their suitability as positive or negative electrodes, and quantitatively assess their electron affinity properties.

\begin{figure}
    \centering
    \includegraphics[width=0.92\linewidth]{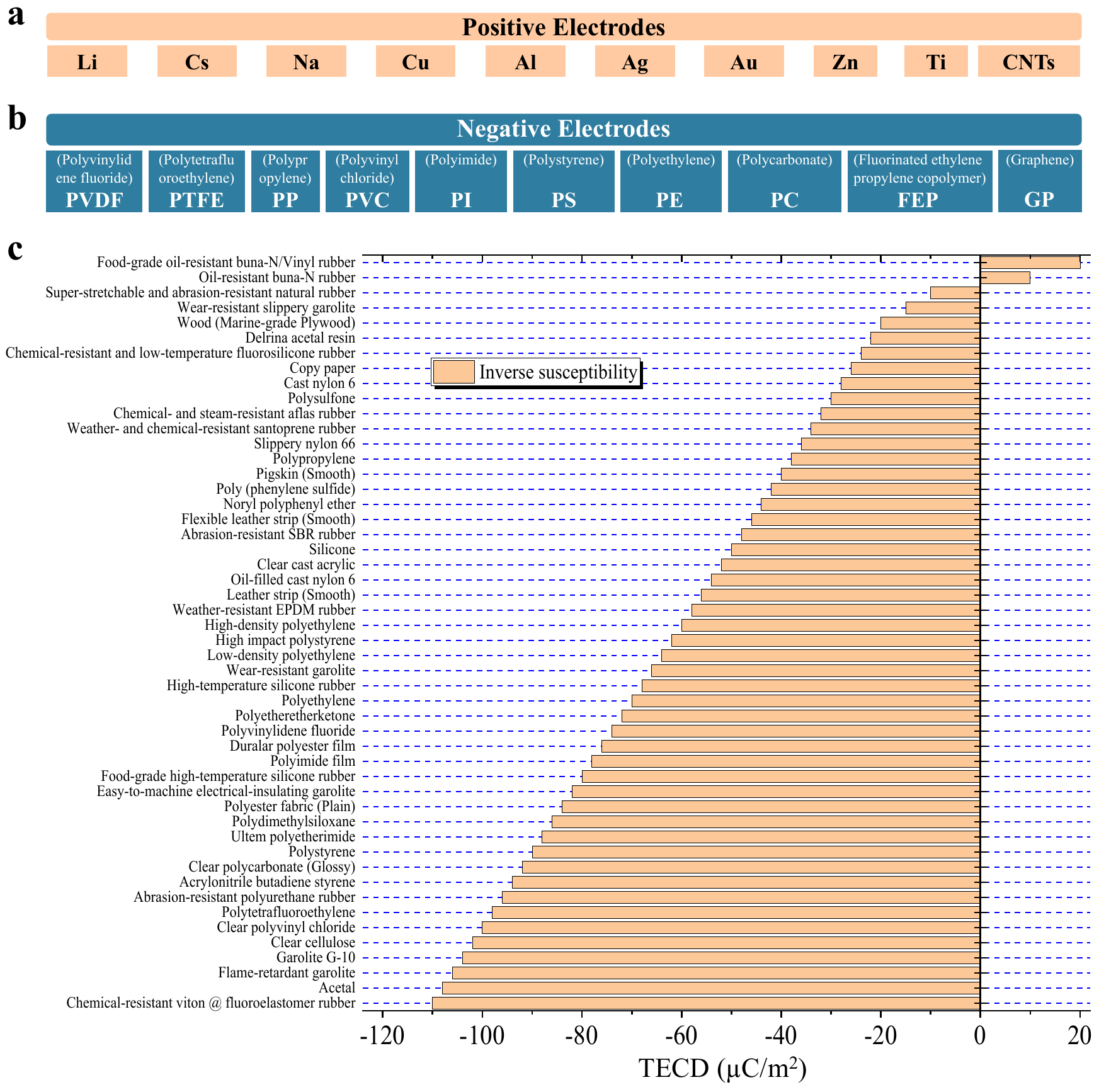}
    \caption{Classification of materials as TENG positive and negative electrodes under unsupervised learning using a convolutional neural network model. (a) Predicted materials suitable for positive electrodes. (b) Predicted materials suitable for negative electrodes. (c) The ability of common materials to gain and lose electrons predicted based on the quantized triboelectric series.}
    \label{Classification}
\end{figure}

The exceptional versatility of TENGs, combined with their environmentally sustainable characteristics and green energy potential, necessitates the development of a robust material screening framework. This framework must effectively classify diverse electrode materials, evaluate their suitability as positive or negative electrodes, and quantitatively assess their electron affinity characteristics. As illustrated in Figs.~\ref{Classification}(a) and (b), our unsupervised learning paradigm demonstrates remarkable classification accuracy through a set of simple instructions. The model autonomously categorizes common materials into positive or negative electrode classifications for TENG devices. The classification results exhibit complete consistency with both experimental observations and established literature, thereby validating the precision and reliability of our unsupervised learning architecture. For example, the model accurately identifies metallic elements (e.g., lithium (Li), copper (Cu), and aluminum (Al)) as appropriate positive electrode materials while correctly classifying polymers such as polytetrafluoroethylene (PTFE), polyimide (PI), and polycarbonate (PC) as negative electrode materials. Moreover, this framework enables a deeper understanding of material properties through systematic classification.

Fig.~\ref{Classification}(c) presents a quantitative ranking of material electron affinity characteristics, demonstrating excellent agreement with published experimental data. The hierarchy of rankings shows strong consistency with the triboelectric series established by Wang \textit{et al.} in 2019~\cite{zou2019quantifying}, thus confirming the physical validity of our predictive model. This material classification and screening model provides a powerful tool for rational electrode material selection in TENG design. Furthermore, the integration of doping materials screening capabilities enables performance-optimized material selection based on predicted output characteristics, representing a significant advancement in TENG material informatics. These advancements highlight the model's potential for broader applications in material science.

The developed model establishes a novel approach for intelligent material selection in triboelectric energy harvesting systems, offering substantial improvements over conventional trial-and-error approaches. The demonstrated accuracy in material classification and prediction of electronic gain and loss capability, coupled with the ability of the model to predict doping materials, provides a comprehensive solution for performance-optimized TENG design. This computational framework not only accelerates material discovery but also enhances our fundamental understanding of structure-property relationships in triboelectric materials.

\subsection{Optimized material screening for TENG electrodes}

To optimize polymer electrode selection in TENGs, we developed a robust ML model that systematically correlates TENG performance metrics with the intrinsic properties of electrode materials. This model rigorously controls material-specific variables, ensuring accurate predictions for polymer-based negative electrodes and enabling comprehensive material screening.

To establish a robust ML model for optimizing negative electrode selection in TENGs, we developed a comprehensive material screening model. By systematically analyzing thousands of peer-reviewed studies, we quantified the fundamental correlations between the performance metrics of TENGs (output current, voltage, power density, and cyclic stability) and the properties of electrode materials. During model development, we rigorously controlled parameters by isolating material-specific variables from external factors such as surface morphology, contact area, and electrode thickness. This approach ensured the model's predictive accuracy for polymer-based negative electrode materials. Based on this model, we screened various negative electrode materials and analyzed their performance characteristics.

As demonstrated in Fig.~\ref{Performance}(a), our material screening reveals that pairing PTFE with positive copper electrodes achieves exceptional performance, producing an output voltage of 325~V and an energy density of 0.35~J/cm$^2$. The predicted data align closely with results reported in the literature and experiments~\cite{qian2023fluoropolymer,zhang2020material, yang2013single}. Notably, our prediction identifies graphene (GP) as an outstanding negative electrode material, achieving superior performance metrics, including a 375~V output voltage and 0.45~J/cm$^2$ energy density. This remarkable performance is attributed to graphene's exceptional surface area characteristics~\cite{worsley2011high}. Despite high output performance, the practical implementation of GP electrodes faces significant challenges, such as high material costs and complex pretreatment requirements.

The predicted performance hierarchy of conventional polymer materials, including polypropylene (PP), polyvinylidene fluoride (PVDF), PI, and polyvinyl chloride (PVC), exhibits remarkable agreement with the established triboelectric series~\cite{zou2019quantifying}. This correlation not only validates the predictive accuracy of our ML model but also provides fundamental insights into the structure-property relationships governing triboelectric performance. Our developed model marks a significant advancement in data-driven material selection for TENG applications, offering a systematic approach to optimize electrode material screening while considering both performance metrics and practical implementation constraints.

\subsection{Optimized doping strategies for TENGs output performance}

To enhance the performance of TENGs, we systematically investigated the impact of material doping strategies on key performance metrics, including output current, voltage, and energy density emphatically. Our computational model identified optimal doping materials and concentrations, providing a framework for performance optimization through enhanced charge capture efficiency and transfer kinetics.

\begin{figure}
    \centering
    \includegraphics[width=0.92\linewidth]{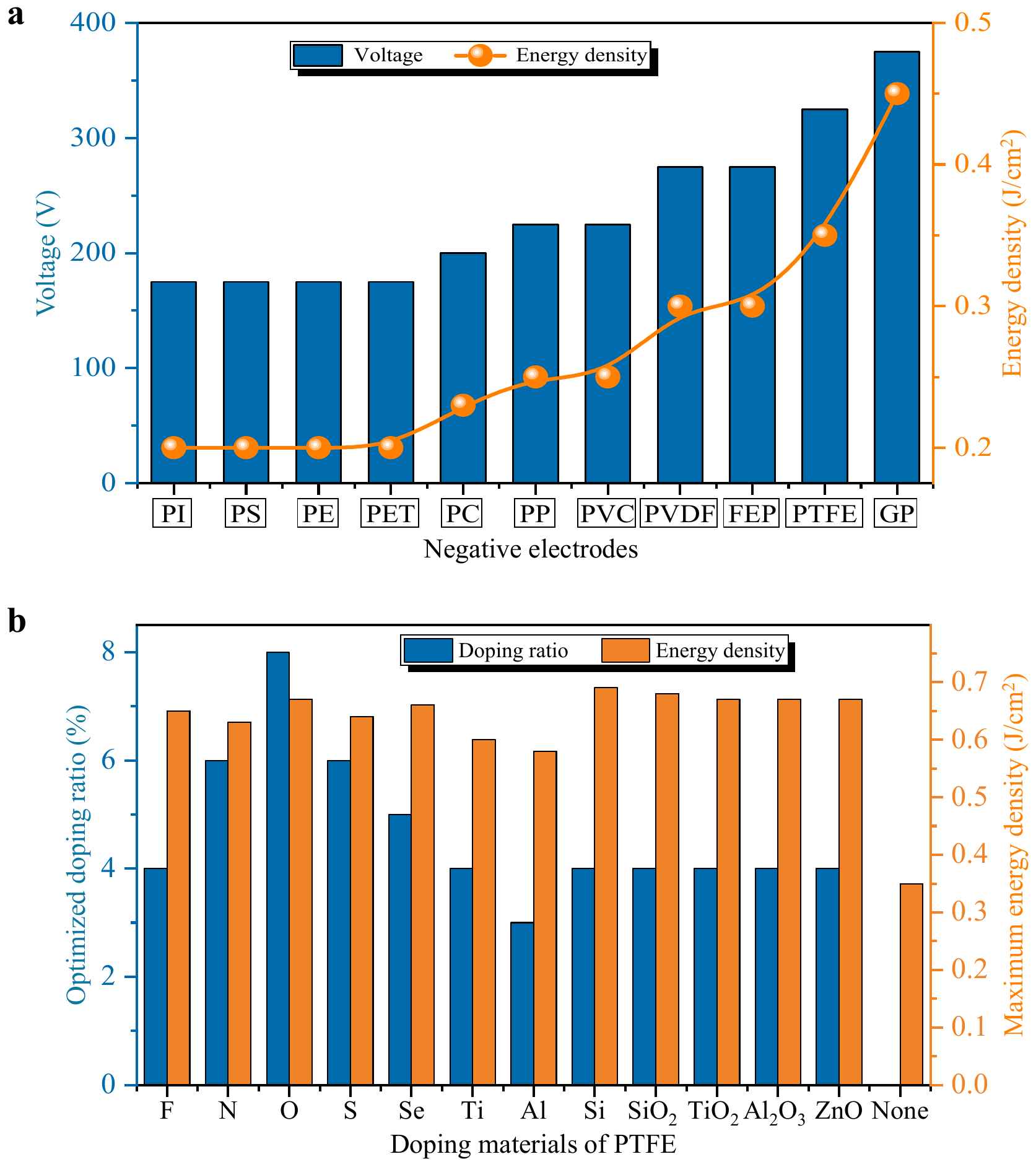}
    \caption{Predicted TENG output performance with different negative electrode materials selected using Cu as the positive electrode. (a) Output voltage (V) and energy density as functions of negative electrode materials. (b) Optimized maximum energy density of TENG performance as a function of doping materials when PTFE is selected as the negative electrode material, corresponding to the optimal doping ratio, as predicted by a convolutional neural network model.}
    \label{Performance}
\end{figure}

\begin{figure}
    \centering
    \includegraphics[width=0.92\linewidth]{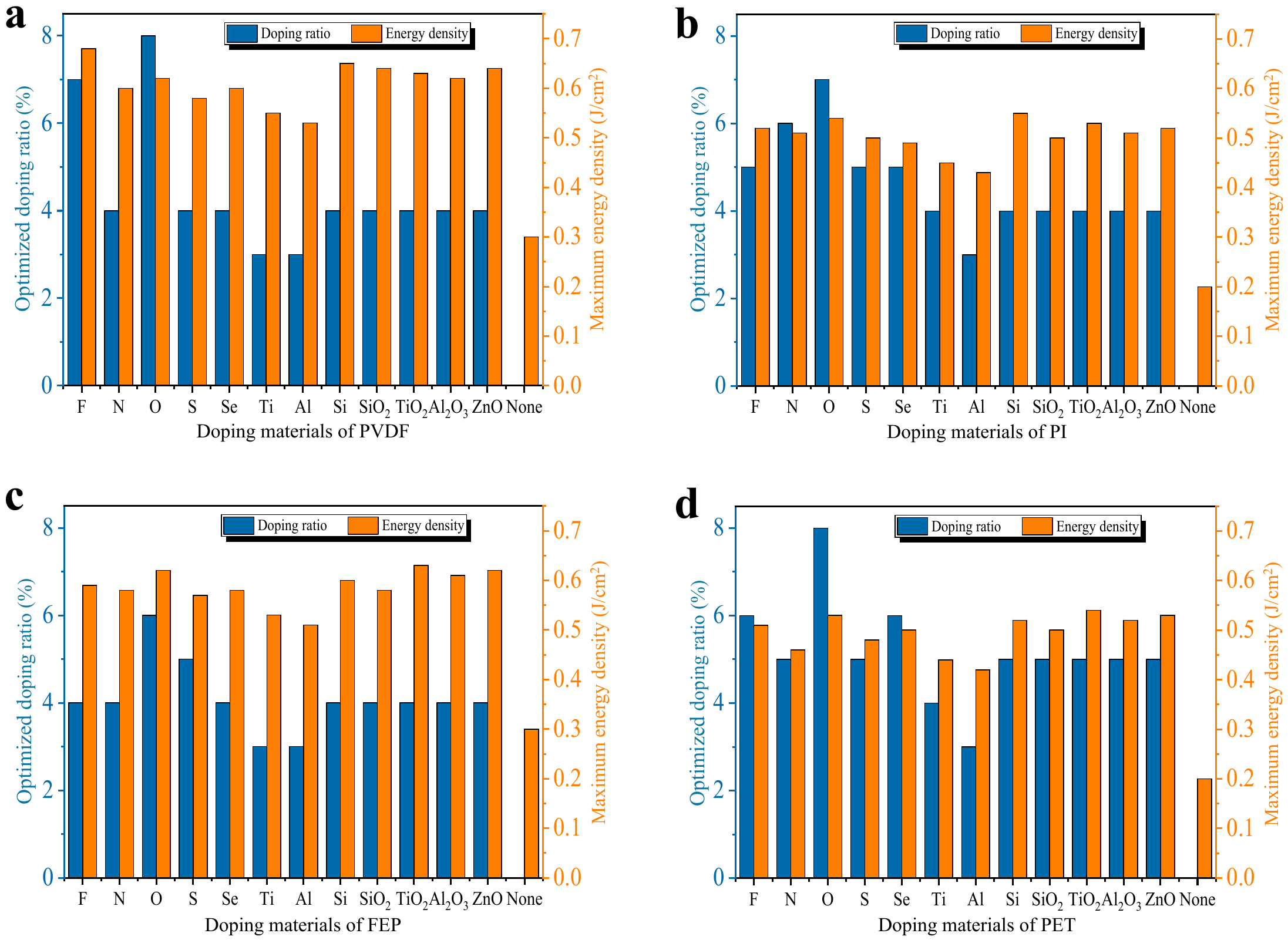}
    \caption{Predicted TENG output performance when Cu is used as the positive electrode and different types of negative electrode materials are selected, with optimal doping ratios for different doping materials. The optimal doping ratios (left vertical axis) as well as the corresponding maximum energy density (right vertical axis) as a function of the selected doping materials for negative electrode materials: (a) PVDF, (b) PI, (c) FEP, and (d) PET, as predicted by a convolutional neural network model.}
    \label{Ratio}
\end{figure}

Following the initial validation of our model's predictive accuracy, we extended our investigation to systematically analyze the performance enhancement of TENGs through various material doping strategies. Our study focused on key performance metrics, including output current, voltage, and energy density. Figs.~\ref{Performance}(b) and~\ref{Ratio} illustrate the comparative performance analysis of diverse polymers enhanced with a range of dopant materials. Based on this model, we explored the influence of different doping strategies on TENG performance.

The computational model identified undoped PTFE as a promising candidate (Fig.~\ref{Performance}(a)), demonstrating substantial output energy density and cost-effectiveness for practical applications. Our analysis revealed that material doping significantly influences TENG performance characteristics (Fig.~\ref{Performance}(b)). Specifically, incorporating elemental dopants (fluorine (F), nitrogen (N), aluminum (Al)) and oxide materials (silicon dioxide (SiO$_2$), titanium dioxide (TiO$_2$)) into PTFE matrices resulted in notable improvements in output voltage and energy density (Fig.~\ref{Performance}(b)). The doping optimization process was guided by our predictive model, which identified optimal doping ratios to maximize performance. As shown in Fig.~\ref{Performance}, introducing Al dopant at 3\% concentration enhanced the energy density from 0.35~J/cm$^2$ (undoped PTFE) to 0.58~J/cm$^2$, increased by $\sim$ 65.7\%. This improvement is attributed to enhanced charge capture efficiency and transfer kinetics. Fluorine-doped PTFE, in particular, exhibited favorable characteristics~\cite{wu2024surface}. In our study, the doping of F at 4\% concentration increased the energy density of PTFE up to 0.65~J/cm$^2$ (Fig.~\ref{Performance}(b)), increased by $\sim$ 85.7\%. Our analysis revealed distinct doping efficiency trends, with certain materials achieving substantial performance gains at low concentrations, while others required higher doping ratios for comparable improvements.

Comparative analysis of oxide-doped systems revealed superior performance enhancement compared to elemental doping, primarily due to improved dielectric properties and enhanced charge generation and transfer mechanisms~\cite{wang2017achieving}. Fig.~\ref{Ratio} illustrates the performance optimization of various polymer matrices (PVDF (Fig.~\ref{Ratio}(a)), PI (Fig.~\ref{Ratio}(b)), FEP (Fig.~\ref{Ratio}(c)), and PET (Fig.~\ref{Ratio}(d))) through different doping materials. Notably, a 4\% TiO$_2$ doping ratio enhanced energy density up to 0.63~J/cm$^2$, 0.53~J/cm$^2$, and 0.63~J/cm$^2$ in PVDF, PI, and FEP, respectively. Even undoped PET, which exhibited relatively poor energy density (0.2~J/cm$^2$) as shown in Fig.~\ref{Ratio}(d), achieved an energy density of 0.54~J/cm$^2$ (enhanced by $\sim$ 170\%) with 5\% TiO$_2$ doping. Interestingly, oxygen doping demonstrated limited performance improvement across various polymer matrices, requiring higher concentrations while yielding suboptimal energy density enhancements. These performance improvements are due to the lower electron affinity of oxygen. These findings align with fundamental triboelectric principles and established physical theories, validating the robustness of our predictive model. The developed model provides valuable insights into structure-property relationships in doped triboelectric materials, offering a systematic approach for performance optimization in TENG applications.

\subsection{Optimization of doping ratios for TENG performance}

We systematically investigated the relationship between doping ratios and TENG output performance. Optimizing the performance of TENG through doping strategies involves understanding the critical roles of doping materials, substrate materials, and doping ratios. By leveraging a predictive model, the optimal doping ratios can be identified to maximize energy density, providing both theoretical and practical frameworks for enhancing TENG output \cite{wang2017achieving-1}.

Building on the success of our predictive model, we further investigated the relationship between the doping ratio of materials applied to the polymer substrates and the resulting output performance. PTFE was selected as the substrate material due to its excellent performance characteristics, while titanium dioxide (TiO$_2$) was chosen as the primary dopant for its demonstrated ability to enhance efficiency and output performance. Based on this foundation, we analyzed how varying doping ratios influence energy density and overall output performance.

\begin{figure}
    \centering
    \includegraphics[width=0.92\linewidth]{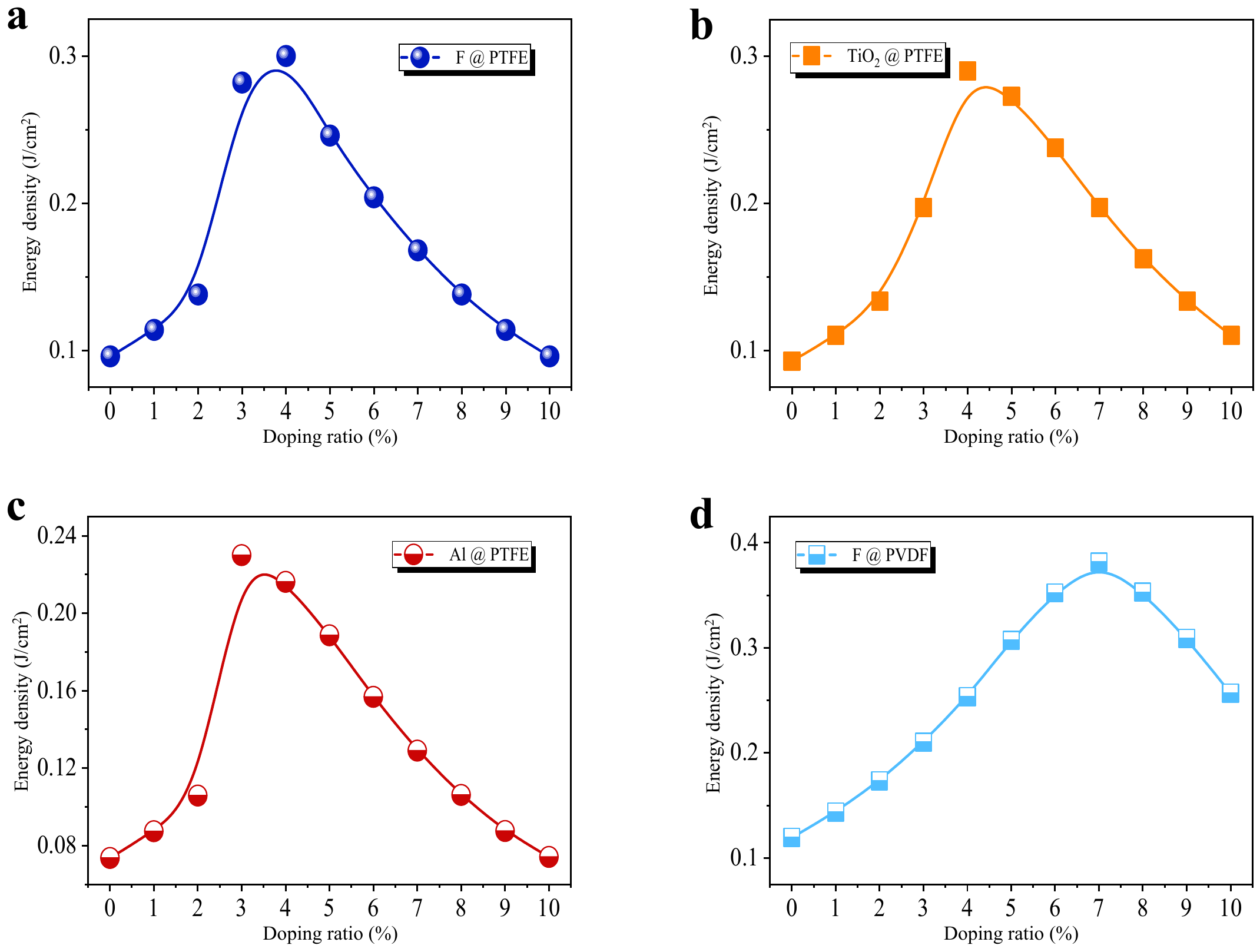}
    \caption{Predicted TENG output performance with different doping materials and doping ratios when Cu is used as the positive electrode. Energy density as a function of doping ratio. Doping materials and substrate materials include: (a) F doping in PTFE, (b) TiO$_2$ doping in PTFE, (c) Al doping in PTFE, and (d) F doping in PVDF, as predicted by a convolutional neural network model.}
    \label{Density}
\end{figure}

As shown in Fig.~\ref{Density}(a), starting with a 0\% doping ratio, the output energy density exhibited a significant increase as the fluorine doping ratio increased. Specifically, when the doping ratio reached 4\%, the energy density was maximized at 0.3~J/cm$^2$. However, further increases in the doping ratio resulted in a decline in energy density. Similarly, doping with other materials in PTFE, such as TiO$_2$ (Fig.~\ref{Density}(b)) and Al (Fig.~\ref{Density}(c)), and F doped in PVDF (Fig.~\ref{Density}(d)) followed a comparable trend: the energy density initially increased before decreasing at higher doping ratios. This trend suggests the existence of an optimal doping ratio for each material, beyond which performance begins to deteriorate.

These findings confirm the existence of an optimal doping ratio that maximizes output performance. This trend aligns with experimental observations, demonstrating that each material has a specific doping ratio corresponding to its maximum output efficiency. Moderate doping enhances charge capture and transfer efficiency, increases charge shielding effects, but compromises material stability, ultimately decreasing energy density.

The objective is to minimize the proportion of doping materials while enhancing performance. By leveraging large-scale model predictions, we can narrow the range of doping ratios, reduce screening time, and efficiently identify the optimal doping ratio. This approach minimizes the influence of doping materials on parent materials, reducing structural and performance interference while lowering manufacturing costs for TENG devices. Additionally, our predictive model provides valuable insights for doping experiments and industrial design, improving material selection processes and enhancing prediction accuracy.

\subsection{Performance optimization of TENG via doping strategies based on physical and chemical mechanisms}

Table~\ref{Table1} presents the doping ratios corresponding to the optimal energy density achieved by incorporating various doping materials into different polymers. By comparing the energy density before and after doping, it is evident that different doping materials exert varying effects on the substrate materials. For instance, certain materials, such as Al, require only a small amount of doping to significantly enhance the output energy density when doped into substrates like PTFE, PVDF, and PS. In contrast, oxygen doping requires a higher doping ratio to achieve substantial improvements in energy density. These findings underscore the critical influence of the doping material, substrate material, and doping ratio on the performance of TENG, necessitating a comprehensive evaluation of these factors.

Doping not only enhances conductivity but also impacts energy density by altering the material's electronic structure. Charge trapping occurs when dopants introduce defects or intermediate states in the material, which captures surface charges \cite{zhang2016surface}, increasing charge density and enhancing energy storage \cite{shuai2014charge}. Additionally, doping can modify the surface charge distribution, affecting charge mobility and trapping dynamics, which optimize energy release. On the other hand, doping also affects the material’s bandgap. Changes in the bandgap directly influence the excitation and transfer of electrons, which in turn alters the energy output \cite{xie2023effective}. Doping elements can shift the bandgap to improve the material's conductivity or create optimal conditions for charge transfer, thus influencing the efficiency of the triboelectric process.

Our ML model, based on GNNs, predicts the optimal doping ratio by integrating structural and electronic data. The model learns how doping influences surface charge density, charge mobility, and bandgap size, finding the ratio that maximizes energy output. Our results show that the doping ratios predicted by the model align with established triboelectric theories, which state that higher surface charge density, enhanced electron mobility, and an optimized bandgap result in improved energy harvesting efficiency. Overall, the optimal doping ratios predicted by our ML model are consistent with existing triboelectric theories, showing that doping affects energy density through mechanisms such as charge trapping and bandgap modulation.

The optimization of doping is guided by our predictive model, which identifies the optimal doping ratio to maximize performance. Table~\ref{Table5} presents the percentage deviations between the model predictions and the experimental results. This predictive model not only provides a theoretical foundation for doping optimization but also serves as a practical tool for evaluating and enhancing TENG output performance in specific systems. For example, the predicted energy density for Ag doped into PTFE, with a Cu negative electrode, is 1.12~J/cm$^2$. This value could serve as a standard for pure Cu doped into pure PTFE. Future experiments can leverage this predicted value as a baseline to analyze enhancements by comparing experimental results with the model’s predictions \cite{Kim2019}.

\subsection{Validation of TENG predictive model}

The predictive model for TENGs was validated through experimental data, demonstrating its ability to classify materials, predict output performance with 98\% accuracy, and evaluate the effects of doping strategies. These results confirm the model's robustness and applicability for optimizing TENG design and performance.

To validate the accuracy and efficiency of the model, we selected eight distinct sets of experimental data from TENG electrodes, none of which were included in the model's training dataset. This approach ensured the feasibility and robustness of the model. The following experimental steps were conducted to evaluate the model's performance:

\begin{enumerate}
    \item First, unclassified TENG electrode materials were input into the model. Under unsupervised conditions, the model automatically classified the positive and negative TENG electrode materials with 100\% accuracy. Additionally, it classified the materials' ability to gain and lose electrons, also achieving 100\% accuracy. This classification capability established a foundation for subsequent performance predictions.
    
    \item Next, the model predicted the output performance of these materials and ranked them. It achieved an accuracy of 98\%, with results closely aligning with the electrostatic sequence table. This high level of accuracy demonstrated the model's reliability in performance prediction.
    
    \item Subsequently, the model was tested for its ability to evaluate the impact of various doping materials on TENG output performance. The results provided valuable insights into selecting effective doping materials to enhance TENG performance, such as identifying specific doping materials and their corresponding effects on output efficiency.
    
    \item Finally, the model was used to predict the output trends of TENG under randomly selected doping ratios. The predictions were highly consistent with the observed experimental trends, further validating the model's reliability and applicability for optimizing doping ratios.
\end{enumerate}

To evaluate the stability and robustness of the model’s predictions, we performed a statistical analysis on the predicted energy densities. Specifically, we used 10-fold Monte Carlo cross-validation to assess the variability in the predicted results. In each iteration, the data were randomly split into training and validation sets, and the model was retrained to predict energy densities for each material-dopant combination.

For each prediction, the mean (\(\mu\)) and standard deviation (\(\sigma\)) of the predicted energy densities were calculated using the following formulas:
\begin{equation}
\mu = \frac{1}{n} \sum_{i=1}^{n} y_i, \textrm{and}
\end{equation}
\begin{equation}
\sigma = \sqrt{\frac{1}{n-1} \sum_{i=1}^{n} (y_i - \mu)^2},
\end{equation}
where \( y_i \) represents the predicted energy density for the \(i\)-th iteration, and \(n\) is the total number of predictions (in our case, \(n = 10\)).

Additionally, the 95\% confidence interval (CI) was calculated to quantify the uncertainty in the predictions. The confidence interval is given by:
\begin{equation}
\textrm{CI} = \mu \pm t_{(0.025, (n-1))} \cdot \frac{\sigma}{\sqrt{n}},
\end{equation}
where $t_{(0.025, (n-1))}$ is the critical value from the $t$-distribution with $(n-1)$ degrees of freedom. This confidence interval provides a range within which we expect the true value of the energy density to fall with 95\% confidence.

Our analysis showed that most of the predicted energy densities exhibited small standard deviations, typically below 0.05 J/cm², indicating that the model’s predictions are stable and reliable. The confidence intervals for these predictions were narrow, reflecting low uncertainty in the model’s output. However, for some material-dopant combinations, such as Ag-doped PTFE and TiO$_2$-doped PVDF, the standard deviations were higher, and the confidence intervals were wider, suggesting that the model’s predictions are more sensitive to dopant concentrations and material properties for these systems.

Potential sources of uncertainty in the model's predictions include noise in the training dataset, which originates from variations in experimental conditions across different studies; the simplifications made in the model, such as the exclusion of factors like surface morphology; and the sensitivity of the model to hyperparameter tuning, particularly during the early stages of training. These factors contribute to the variability in the predictions and are addressed in ongoing work, where we aim to apply Bayesian neural networks to better quantify and incorporate prediction uncertainty.

Despite these sources of uncertainty, the findings highlight the accuracy and efficiency of the model in classifying materials, predicting performance, and providing insights into doping strategies. The validated model offers a robust framework for advancing TENG design and optimization, with significant implications for both experimental research and industrial applications.

\section{Conclusions}

In this study, we developed a robust ML-driven model to optimize the performance of TENGs by leveraging GNNs. This approach systematically integrates experimental and computational data to predict electrode material properties, identify optimal doping strategies, and establish structure-property relationships. The model demonstrated exceptional accuracy in material classification and performance prediction, achieving up to 0.83 J/cm² energy density with 4\% SiO$_2$-doped PTFE paired with a Cu electrode. Key findings include a 65.7\% energy density enhancement for aluminum-doped PTFE and an 85.7\% improvement for fluorine-doped PTFE, highlighting the critical role of doping materials and concentrations. Based on these findings, we proposed a framework that accelerates material discovery, reduces experimental costs, and provides novel insights into the fundamental determinants of TENG performance. Additionally, the integration of advanced GNN methodologies enables high-throughput screening of material combinations, offering a transformative approach to intelligent material design. Despite its advantages, the study is limited by reliance on available experimental data for model validation and challenges in data interpretability. Future work will focus on refining the GNN model, validating additional materials experimentally, and applying the approach to real-world TENG devices. This work not only advances the field of sustainable energy harvesting but also establishes a foundation for the development of next-generation self-powered systems and environmentally friendly energy technologies.

\clearpage

\section{Experimental Section}

\subsection{Machine learning techniques and their application in TENGs}

ML, a subfield of artificial intelligence, enables machines to learn from data and improve performance over time without explicit programming. ML algorithms identify patterns in data to make predictions or decisions, eliminating the need for task-specific programming \cite{yu2019dag}. ML can be broadly categorized into three types: supervised learning \cite{ZHAO2024112982, Zhao2024-1, 63, Zhijing2025}, unsupervised learning \cite{BO2025111468, Liu2025}, and reinforcement learning \cite{10675394, LIU2025110279}, each suited to specific application scenarios. Based on these frameworks, ML has become a focus of research in optimizing TENG performance.

In supervised learning, models are trained on labeled data, where both input features and corresponding outputs are provided. The objective is to learn a mapping from inputs to outputs, which can subsequently be used to predict outputs for unseen data \cite{dai2021graph}. For instance, in regression tasks, the aim is to find a function that predicts a continuous output based on input features, whereas in classification tasks, the goal is to assign data to predefined categories \cite{fung2021benchmarking, reiser2022graph}. The supervised learning model can be mathematically expressed as:
\begin{equation}
    y = f(x; \theta),
    \label{1}
\end{equation}
where \(x\) represents the input features, \(y\) is the output, and \(\theta\) denotes the parameters learned from the training data \cite{dong2023sli}. This model is widely utilized in predictive modeling tasks, including the prediction of material properties. In TENG applications, supervised learning can be employed to predict material properties such as energy density and mechanical strain response.

Unsupervised learning, in contrast, deals with unlabeled data. The goal is to uncover hidden structures or patterns within the data \cite{xie2016unsupervised}. Common techniques include clustering, which groups data points based on similarity, and dimensionality reduction, which reduces the number of features in a dataset while preserving critical information. The clustering algorithm can be defined as:
\begin{equation}
    J = \sum_{i=1}^{n} \|x_i - \mu_{k_i}\|^2,
    \label{2}
\end{equation}
where \(x_i\) is a data point, \(\mu_{k_i}\) represents the centroid of cluster \(k_i\), and \(J\) is the cost function to be minimized. Minimizing \(J\) yields optimal clustering results. In TENG applications, unsupervised learning can identify hidden patterns in material performance, guiding material selection.

Reinforcement learning is another paradigm in which an agent learns to make decisions by interacting with its environment. The agent receives feedback in the form of rewards or penalties based on its actions, enabling it to adapt its strategy over time. The value function \(V\) in reinforcement learning is expressed as:
\begin{equation}
    V(s) = \mathbb{E} \left[ \sum_{t=0}^{\infty} \gamma^t R_t \right],
    \label{3}
\end{equation}
where \(V(s)\) is the value of being in state \(s\), \(R_t\) is the reward received at time step \(t\), and \(\gamma\) is the discount factor that determines the weight of future rewards. In TENG optimization, reinforcement learning could be applied to optimize dynamic operating conditions by simulating environmental feedback to enhance energy collection efficiency.

\begin{figure}
    \centering
    \includegraphics[width=0.82\linewidth]{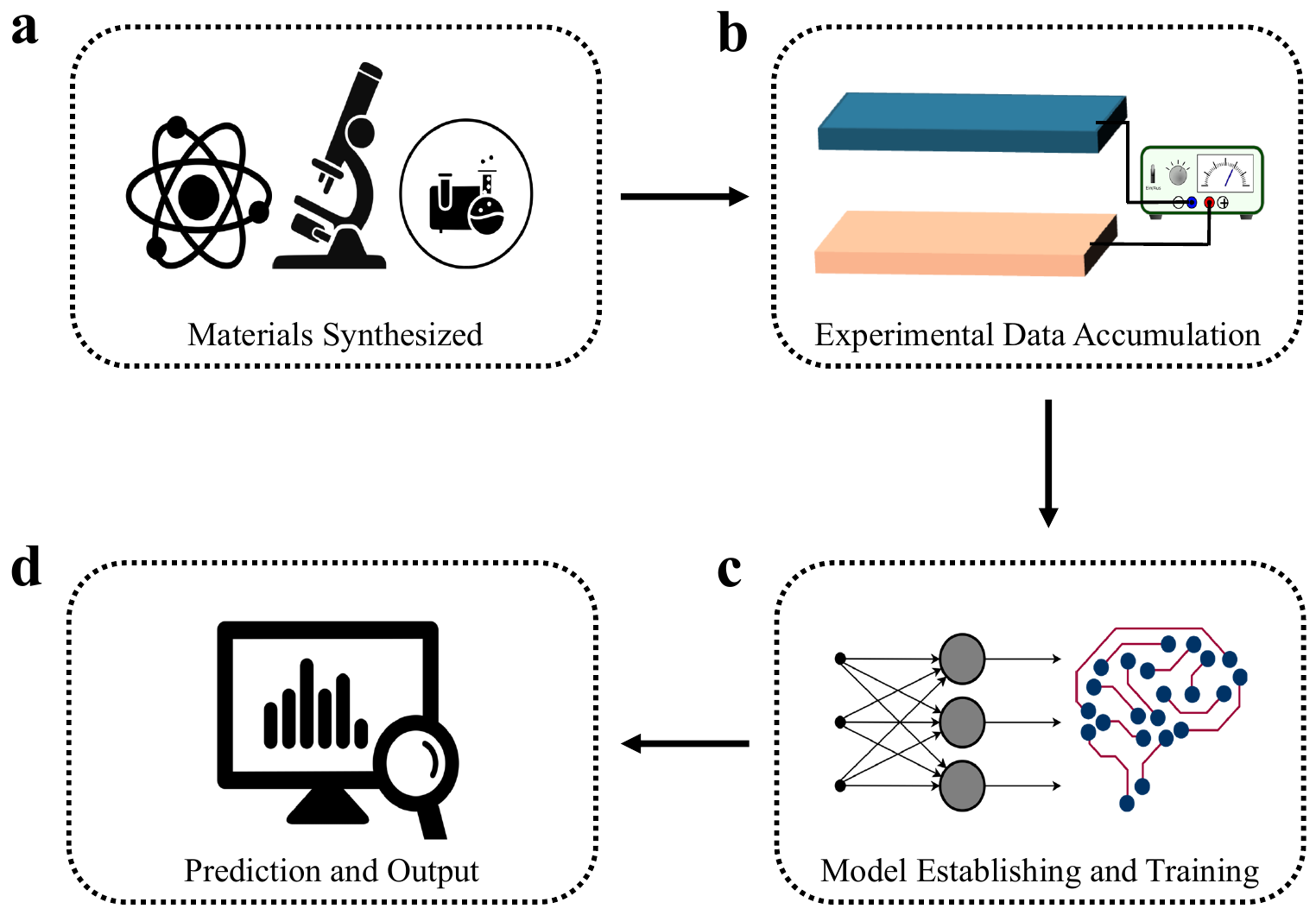}
    \caption{Workflow for predicting the output performance of the negative electrodes in TENG. (a) Synthesized materials were extracted from literature. (b) Experimental data were collected from prior studies. (c) The data were integrated into a convolutional neural network model for training. (d) The output performance and accuracy were evaluated using the predicted results.}
    \label{Workflow}
\end{figure}

In the context of TENGs, ML techniques are primarily utilized to predict energy density, a critical parameter for optimizing device performance and ensuring efficient energy harvesting. Fig.~\ref{Workflow} illustrates a structured workflow for implementing ML in this domain, focusing on the prediction of negative electrode output performance. The process begins with material synthesis (Fig.~\ref{Workflow}(a)), where candidate electrode materials are selected based on prior studies. Subsequently, the experimental data are systematically collected and analyzed (Fig.~\ref{Workflow}(b)) to construct a comprehensive dataset for model training. This dataset is then integrated into a CNN model (Fig.~\ref{Workflow}(c)), which is trained to identify correlations between material properties and output performance. Finally, the trained model predicts performance metrics, and these predictions are validated against experimental results (Fig.~\ref{Workflow}(d)). This data-driven workflow streamlines the optimization of electrode materials, enhancing the predictive power of ML in energy harvesting research.

ML models, such as regression or neural networks, can be trained on experimental data, where inputs include material properties, mechanical stresses, and other operational conditions, and the target output is the predicted energy density. Such models can be represented as:
\begin{equation}
    E_d = f(x_1, x_2, ..., x_n; \theta),
    \label{4}
\end{equation}
where \(f\) is the learned function, \(x_1, x_2, ..., x_n\) denote the input features (e.g., material properties, stress levels), and \(\theta\) represents the parameters learned during training. This approach facilitates the optimization of TENG design and operation, thereby enhancing energy harvesting efficiency under varying conditions.

\subsection{Data collection and machine learning applications in TENGs}

For ML algorithms to be effective in TENGs, it is crucial to gather accurate and comprehensive data. The primary types of data include voltage, current, and energy density, which are essential for constructing predictive models aimed at optimizing TENG performance. These measurements are typically obtained through sensors that capture the electrical output and mechanical behavior of the device. In addition to experimental data, open-access literature serves as a complementary source of information, further enriching ML models. To achieve efficient data collection, we developed an automated approach that extracts relevant information from publicly available literature.

By employing Python-based scripts, we batch-scraped and downloaded relevant publications from public literature databases, resulting in a dataset comprising approximately 6,000 data points. This automated approach not only improved data collection efficiency but also enhanced dataset diversity, enabling more robust ML analysis. These datasets supported the analysis of key variables such as voltage, current, and energy density.

The voltage ($V$) generated by a TENG arises directly from the triboelectric effect and the mechanical stress applied to the materials. It is influenced by factors such as material selection, device geometry, and the rate of mechanical motion. Monitoring voltage allows researchers to evaluate the electrical output of the device under varying conditions, thereby refining TENG design and operation. In addition to voltage, current is another critical variable that reflects the energy generated by the TENG.

Similarly, the current ($I$) is another critical variable that provides insights into the energy generated by the TENG. Current is proportional to the rate of charge flow induced by the triboelectric effect. Measuring the current enables researchers to assess device efficiency and monitor performance over time. For example, in self-powered sensors, stable current output is crucial for reliable data transmission.

In addition to voltage and current, energy density, \(E_d\), is a key metric collected during experiments. The energy density in TENGs can be calculated from the measured voltage and capacitance using the formula:
\begin{equation}
    E_d = \frac{1}{2} C V^2,
    \label{5}
\end{equation}
where \(C\) is the capacitance and \(V\) is the voltage \cite{zou2019quantifying}. By collecting energy density data under various operational conditions, researchers can predict the energy output of TENGs across different scenarios. These measurements collectively form the core dataset for training ML models, enabling the prediction of optimal energy harvesting conditions.

The dataset derived from these measurements is essential for training ML models. It enables the models to identify intricate relationships among input variables, such as material properties, mechanical stresses, and environmental conditions, and their effects on energy output. However, data collection presented several challenges, particularly in real-world conditions where factors such as noise, environmental fluctuations, and sensor calibration discrepancies can compromise measurement accuracy. Despite these hurdles, advancements in sensor technology and data acquisition techniques are progressively improving the quality and reliability of the data gathered for TENG systems. For instance, improved temperature compensation in sensors has reduced the impact of environmental fluctuations on data accuracy. These improvements enable more robust model training, thereby enhancing the predictive capabilities of ML algorithms in this field.

\subsection{Machine learning models for TENG optimization}

In our work, we developed two distinct ML models to optimize the performance of TENGs. These models include an unsupervised learning model and a GNN model, each targeting different aspects of TENG performance. Together, these models provided complementary approaches for improving material selection and predicting energy output. The following sections detail the implementation and application of these models.

\subsubsection{Unsupervised learning model: deep embedded clustering}

The unsupervised model utilized the deep embedded clustering (DEC) method, which combines deep learning and clustering in an unsupervised learning model. This approach enables the identification of underlying patterns in the data without requiring prior labels \cite{guo2018deep}. The primary goal of this model was to classify materials based on their suitability as positive or negative electrodes in TENGs.

DEC operates by jointly learning feature representations and clustering assignments. Its objective is to minimize the distance between similar data points while maximizing the separation between different clusters. Mathematically, this is expressed as:
\begin{equation}
\mathcal{L}_{DEC} = \mathcal{L}_{reconstruction} + \alpha \cdot \mathcal{L}_{cluster},
\label{6}
\end{equation}
where \(\mathcal{L}_{reconstruction}\) represents the reconstruction loss used to learn feature representations, \(\mathcal{L}_{cluster}\) is the clustering loss that ensures samples within the same cluster are grouped closely, and \(\alpha\) is a regularization parameter balancing the two losses.

The clustering loss function, \(\mathcal{L}_{cluster}\), is defined as:
\begin{equation}
\mathcal{L}_{cluster} = -\sum_{i=1}^{N} \log \left( \frac{e^{s_{k_i}(x_i)}}{\sum_{k=1}^K e^{s_k(x_i)}} \right),
\label{7}
\end{equation}
where \(s_{k_i}(x_i)\) represents the score assigned to cluster \(k_i\) for sample \(x_i\), and \(K\) is the total number of clusters. This loss function encourages the model to maximize the score for the correct cluster while minimizing the distance between similar data points \cite{guo2017improved}.

Applying this method to our material dataset, the model effectively classified materials into two groups: those suitable as positive electrodes and those better suited as negative electrodes, based on their inherent properties. This classification provided a foundation for subsequent energy density prediction.

With the increasing complexity of material classification problems, traditional clustering algorithms often struggle with high-dimensional datasets, necessitating the exploration of more advanced methods. To further demonstrate the effectiveness of the DEC method for material classification, we conducted a comparative analysis between DEC and two traditional clustering algorithms, namely K-means and density-based spatial clustering of application with noise (DBSCAN). The comparison was based on several performance metrics, including classification accuracy, silhouette score, and computational complexity. These comparative results are summarized in Table~\ref{Table2Comp}. We evaluated the clustering performance of each method using the same material-dopant dataset. The results revealed that DEC significantly outperforms both K-means and DBSCAN, achieving higher classification accuracy and a better-defined clustering structure. Specifically, DEC achieved a higher silhouette score, reflecting the compactness and separation of the clusters, while K-means and DBSCAN struggled with the high-dimensional and complex nature of the material data.

These results highlight the distinct advantages of DEC in material classification tasks, which can be attributed to its ability to learn feature representations during the clustering process. This enables DEC to capture complex, non-linear relationships between material properties and performance metrics. In contrast, K-means and DBSCAN are limited in their ability to model such intricate relationships, particularly in high-dimensional feature spaces. This makes DEC more suitable for handling the intricate and varied characteristics of the material-dopant combinations in our dataset. 

\subsubsection{Graph neural network model: energy density prediction}

\begin{figure}
\centering
\includegraphics[width=0.82\linewidth]{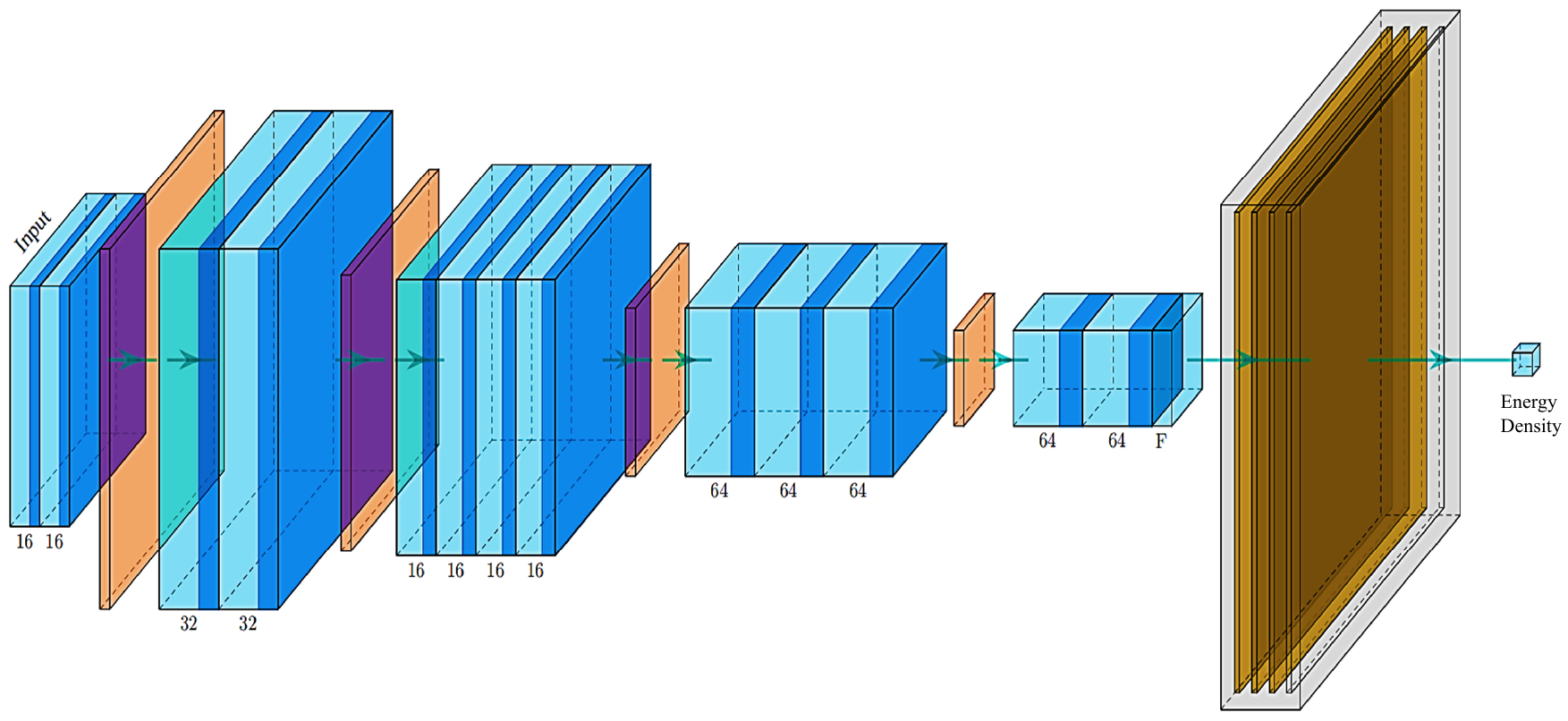}
\caption{Schematic of a four-layer GNN with hierarchical pooling for TENG electrode polarity prediction. The model consists of four attention-based pooling layers that aggregate multiscale material representations. A seven-layer multilayer perceptron translates the pooled features into binary electrode polarity classification, with Batch Normalization and stochastic dropout applied between layers to improve generalization and mitigate overfitting.}
\label{Layer}
\end{figure}

The second model employed is a GNN, which is particularly effective for modeling relational data. As depicted in Fig.~\ref{Layer}, the GNN architecture comprises four graph convolutional layers, four attention-based pooling layers, and seven fully connected layers. To improve generalization and mitigate overfitting, batch normalization and random dropout techniques are applied between layers. GNNs process graph-structured inputs, where nodes represent materials and edges encode relationships between them, defined by their physical properties or interactions within the TENG system. This architecture enables the efficient modeling of complex material interactions, providing a robust model for predicting energy density and optimizing TENG performance.

In our approach, the choice of GNN architecture, particularly the number of layers, was guided by the nature of the dataset and the underlying relationships between material-dopant combinations. We believe the data inherently exhibits clear two-dimensional characteristics due to the interactions between the elements in the dataset, which makes the graph structure particularly suitable for capturing these relationships. GNNs excel in modeling the connections between nodes and edges, allowing the network to learn feature representations that better describe the performance characteristics of specific materials.

The primary objective of the GNN model is to predict the energy density (\(E_d\), as defined in equation~\ref{5}) of various materials when integrated into a TENG. The GNN predicts energy density by aggregating information from neighboring materials (nodes) and updating the node features based on both material properties and interactions.

The core of the GNN is the message-passing mechanism, which enables each node (material) to aggregate information from its neighbors and update its state. The update rule is written as:
\begin{equation}
h_i^{(k+1)} = \sigma \left( W^{(k)} h_i^{(k)} + \sum_{j \in \mathcal{N}(i)} \frac{1}{c_{ij}} W^{(k)} h_j^{(k)} \right),
\label{8}
\end{equation}
where \(h_i^{(k+1)}\) is the updated feature for node \(i\), \(h_i^{(k)}\) is the current feature, \(\mathcal{N}(i)\) denotes the neighbors of node \(i\), \(W^{(k)}\) is the weight matrix for the \(k\)-th layer, \(\sigma\) is a nonlinear activation function (e.g., ReLU), and \(c_{ij}\) is a normalization factor for the edges between nodes \(i\) and \(j\).

The GNN model is particularly effective in capturing complex relationships between materials, such as the effects of material combinations on energy output, which are challenging to model using traditional ML methods. By training the GNN on a large dataset of material properties and corresponding energy densities, we can predict the energy density of new materials that have not been directly measured.

Together, these two ML models—DEC for material classification and GNN for energy density prediction—offer powerful tools for optimizing TENG performance by improving material selection and predicting energy output under varying conditions \cite{dong2023sli}.

\subsection{Training and evaluation of DEC and GNN for TENG optimization}

The training process for the DEC model and the GNN model was conducted using a large dataset comprising experimental data and data sourced from open-access literature. The dataset was carefully preprocessed to ensure consistency and eliminate outliers, enabling the models to learn from high-quality, representative data. These two models, targeting material classification and energy density prediction, provided complementary approaches for optimizing TENG performance. The following sections describe the training processes and evaluation results for both models.

\subsubsection{Training the DEC model}

The DEC model was trained using an unsupervised approach, aiming to optimize feature representations and clustering assignments simultaneously. The training process employed reconstruction loss and clustering loss, with a regularization parameter \(\alpha\) fine-tuned through cross-validation to achieve optimal performance. The model was trained with a batch size of 32 using the Adam optimizer, starting with an initial learning rate of \(10^{-3}\), which was subsequently reduced based on validation loss.

During training, clustering accuracy was evaluated by comparing the predicted labels with the ground truth labels (i.e., whether the materials were categorized as suitable for the positive or negative electrode). After several epochs, the model successfully divided the materials into two distinct groups, demonstrating its ability to learn from unlabeled data and make meaningful classifications based on material properties. This highlights the DEC model's capability as a fast and effective tool for identifying suitable materials for TENG electrodes.

\subsubsection{Training the GNN model}

The GNN model was trained to predict the energy density of materials used in TENGs. Graph-structured data was utilized, where nodes represented materials and edges encoded relationships between materials. The features of each node included various material properties, such as mechanical strength, triboelectric properties, and capacitance.

The training objective was to perform node classification with the energy density \(E_d\) as the target variable. The model employed a graph convolutional network (GCN) layer, and the loss function was defined as the mean squared error (MSE) between the predicted and true energy density values:
\begin{equation}
\mathcal{L}_{GNN} = \frac{1}{N} \sum_{i=1}^{N} \left( E_{d_{pred}}^{(i)} - E_{d_{true}}^{(i)} \right)^2,
\label{9}
\end{equation}
where \(E_{d_{pred}}\) and \(E_{d_{true}}\) are the predicted and true energy densities for the \(i\)-th material, respectively. The Adam optimizer was used for training with a learning rate of \(10^{-4}\) and a batch size of 64.

The training demonstrated that the GNN model effectively learned the complex relationships between material properties and energy density. After 100 epochs, the model’s performance was evaluated using the root MSE (RMSE), which indicated that the GNN could accurately predict energy densities, even for materials that had not been directly measured. This result underscores the GNN model's potential for guiding material design in TENG systems.

\subsubsection{Evaluation and performance}

The performance of both models was rigorously assessed through cross-validation and evaluation on an independent test set. The DEC model achieved a high classification accuracy of 89\% in distinguishing between positive and negative electrode materials, significantly outperforming conventional clustering methods. Additionally, the GNN model achieved an RMSE of 0.12 in predicting energy density, highlighting its effectiveness in accurately forecasting energy density values across a diverse range of materials.

Overall, the results confirmed the high effectiveness of both the DEC and GNN models in optimizing TENG performance. These models provided valuable insights into material selection and serve as reliable tools for designing more efficient TENG systems by accurately predicting energy output under various operational conditions.

With the increasing complexity of material classification problems, traditional ML algorithms often struggle to capture intricate, non-linear relationships in high-dimensional datasets. This necessitates the exploration of advanced models like GNNs. To further validate the effectiveness of our GNN-based approach, we conducted a comparative study with several traditional ML algorithms, including Random Forests, Gradient Boosting Trees (GBT), and Feedforward Neural Networks (FNN). The comparison focused on key performance metrics, such as prediction accuracy, computational complexity, and model interpretability. These comparative results are summarized in Table~\ref{Table3Comp}, providing a clear overview of the performance differences among the models. All models were trained on the same material-dopant dataset. The results showed that the GNN outperformed traditional methods, achieving a higher R-squared value of 0.98 compared to 0.85 for Random Forests, 0.88 for GBT, and 0.81 for FNN. This demonstrates the GNN's superior predictive accuracy, particularly in capturing complex, non-linear relationships between material properties and performance metrics, which simpler models like FNN struggled to model effectively.

Despite its superior predictive accuracy, the GNN exhibited higher computational complexity and longer prediction times. The prediction time for the GNN was 50 seconds, compared to 12 seconds for Random Forests, 25 seconds for GBT, and 30 seconds for FNN. This highlights a trade-off between accuracy and computational cost. Additionally, while the GNN excels in predictive performance, its interpretability is moderate compared to Random Forests and FNN, which offer greater ease of interpretation. 

The results from this comparative analysis demonstrate that the GNN-based model is particularly well-suited for capturing complex relationships in high-dimensional feature spaces, offering a notable improvement in prediction accuracy over traditional methods. However, the increased computational time and moderate interpretability highlight areas where traditional models may still offer advantages, especially in scenarios prioritizing model simplicity and interpretability.

To ensure the reliability and robustness of the GNN model's performance in material classification and doping optimization, we performed a detailed dataset partitioning and cross-validation procedure. The dataset was randomly split into training, validation, and test sets with the following proportions: 70\% for training, 15\% for validation, and 15\% for testing (Table~\ref{Table4Data}). The training set was used to train the model, the validation set was employed for hyperparameter tuning and to prevent overfitting, and the test set was used to evaluate the final performance of the model on unseen data.

Furthermore, to enhance the robustness of our results, we implemented k-fold cross-validation (with k = 5) during the training phase. The dataset was randomly divided into five equal parts, and the model was trained and validated on different subsets of the data, with each fold serving as the validation set once. The results were averaged across all folds to provide a more reliable estimate of the model's generalizability and performance. This procedure is crucial for assessing the stability of the model across different subsets of data and ensuring that the reported performance is not due to overfitting on a specific data split.

Table~\ref{Table4Data} summarizes the dataset partitioning and cross-validation procedure, along with the performance of the model on each corresponding dataset. The GNN model achieved excellent performance across all subsets of the dataset, with an R-squared value of 0.98 on the training set, 0.92 on the validation set, and 0.90 on the test set. These results demonstrate the model's robustness and generalizability, as it performed consistently well across different data splits. The slight decrease in performance from the training set to the test set is typical and indicates the model's ability to generalize effectively to unseen data.

In our study, although experimental validation focused primarily on the optimal material/doping system, all model predictions across various doping ratios and materials were conducted under rigorously standardized conditions to ensure fairness and robustness. Specifically, during both simulation and experimental validation, we maintained constant geometric dimensions (electrode area: 5 × 5 cm\textsuperscript{2}), friction mode (contact–separation at 2.5 Hz), and environmental parameters (5\% relative humidity, 25 $^\circ$C). These controlled settings were deliberately selected to eliminate variability from non-material-related factors, thereby isolating the influence of material composition and doping ratio on device performance. This approach provided a consistent basis for all subsequent comparative analyses.

Building on these standardized conditions, when referring to the optimal material/doping system (for example, 7\% Ag-doped PTFE), we highlight the optimized combination of base material and dopant, as identified by our predictive model. All comparative experiments, including prototype device fabrication and key performance tests, were performed under identical conditions to enable a fair and accurate assessment of the effects of doping.

We acknowledge that broader comparative studies under varied conditions could provide additional insights. However, in this work, our primary objective was to isolate and elucidate the role of doping as the principal variable, thereby enabling a clearer interpretation of its influence on TENG performance. We believe that this focused approach enhances the reliability and scientific value of our findings.

\clearpage

\section*{CRediT authorship contribution statement}

\textbf{Guanping Xu:} Conceptualization, data curation, formal analysis, investigation, visualization, writing-original draft. 
\textbf{Zirui Zhao:} Conceptualization, data curation, formal analysis, investigation, visualization, writing-original draft. 
\textbf{Zhong Lin Wang:} Conceptualization, methodology, supervision, visualization, writing-review \& editing.
\textbf{Hai-Feng Li:} Conceptualization, funding acquisition, methodology, project administration, supervision, visualization, writing-review \& editing.

\section*{Declaration of Competing Interest}

The authors declare that they have no known competing financial interests or personal relationships that could have appeared to influence the work reported in this paper.

\section*{Acknowledgments}

This work was supported by the Science and Technology Development Fund, Macao SAR (File Nos. 0090/2021/A2, 0104/2024/AFJ, and 0115/2024/RIB2), University of Macau (MYRG-GRG2024-00158-IAPME), and the Guangdong{-}Hong Kong{-}Macao Joint Laboratory for Neutron Scattering Science and Technology (Grant No. 2019B121205003). 

\section*{Data Availability}

Data will be made available on request.

\clearpage

\bibliographystyle{elsarticle-num-names}
\bibliography{Guanping3}

\clearpage

\clearpage

\clearpage

\clearpage

\clearpage

\begin{table}[!h]
\renewcommand*{\thetable}{\Roman{table}}
\caption{Selected negative electrodes, including PTFE, FEP, PVDF, polyethylene glycol terephthalate (PET), PS, PI, polyether-ether-ketone (PEEK), and polyurethane (PU), were subjected to a doping optimization process. These electrodes were doped with materials such as Ag and Cu to determine the optimal doping ratio (\%) and the corresponding maximum energy density (J/cm$^2$), as calculated using a convolutional neural network model. Cu was used as the positive electrode. For example, doping Ag into PTFE achieves an optimized maximum energy density of 1.12 J/cm$^2$ at an optimal doping ratio of 7\%.}
\label{Table1}
\setlength{\tabcolsep}{3.28mm}{}
\renewcommand{\arraystretch}{1.1}
\begin{tabular}{l|llllllll}
\hline
Doping               & \multicolumn{8}{c}{Optimized maximum energy density (J/cm$^2$) \text{@} the optimal doping ratio (\%)}                                                                                                      \\
materials            & PTFE                  & FEP                    & PVDF                  & PET                    & PS                     & PI                    & PEEK                   & PU                              \\
\hline
Ag                   &  1.12\text{@}7        &  0.63\text{@}3         & 0.86\text{@}5         &  0.82\text{@}7         &  0.76\text{@}5         & 0.53\text{@}5         &  0.86\text{@}5         &  0.86\text{@}5                  \\  
Cu                   &  0.96\text{@}5        &  0.64\text{@}3         & 0.79\text{@}6         &  0.75\text{@}4         &  0.84\text{@}4         & 0.53\text{@}5         &  0.94\text{@}8         &  0.75\text{@}4                  \\
SiO$_2$              &  0.83\text{@}4        &  0.58\text{@}4         & 0.64\text{@}4         &  0.50\text{@}6         &  0.51\text{@}5         & 0.51\text{@}4         &  0.78\text{@}4         &  0.73\text{@}4                  \\
F                    &  0.83\text{@}5        &  0.59\text{@}4         & 0.68\text{@}7         &  0.51\text{@}6         &  0.51\text{@}3         & 0.52\text{@}5         &  0.72\text{@}5         &  0.78\text{@}5                  \\
ZnO                  &  0.83\text{@}5        &  0.62\text{@}4         & 0.64\text{@}4         &  0.53\text{@}5         &  0.53\text{@}4         & 0.52\text{@}4         &  0.69\text{@}4         &  0.82\text{@}5                  \\
Al                   &  0.83\text{@}4        &  0.51\text{@}3         & 0.53\text{@}3         &  0.42\text{@}3         &  0.61\text{@}5         & 0.45\text{@}4         &  0.71\text{@}4         &  0.73\text{@}5                  \\
TiO$_2$              &  0.76\text{@}5        &  0.63\text{@}4         & 0.63\text{@}4         &  0.54\text{@}5         &  0.48\text{@}3         & 0.53\text{@}4         &  0.74\text{@}5         &  0.73\text{@}5                  \\
Zn                   &  0.76\text{@}4        &  0.55\text{@}3         & 0.59\text{@}5         &  0.48\text{@}5         &  0.51\text{@}4         & 0.47\text{@}3         &  0.64\text{@}3         &  0.73\text{@}4                  \\
Si                   &  0.76\text{@}4        &  0.51\text{@}4         & 0.43\text{@}4         &  0.51\text{@}3         &  0.48\text{@}5         & 0.52\text{@}5         &  0.64\text{@}5         &  0.73\text{@}3                  \\
Ti                   &  0.45\text{@}6        &  0.62\text{@}5         & 0.48\text{@}5         &  0.49\text{@}4         &  0.61\text{@}4         & 0.53\text{@}4         &  0.71\text{@}3         &  0.73\text{@}4                  \\
\hline 
\end{tabular}
\end{table}

\clearpage

\begin{table}[!ht]
\centering
\renewcommand*{\thetable}{\Roman{table}}
\caption{Performance comparison of clustering algorithms (DEC, K-means, and DBSCAN).}
\label{Table2Comp}
\setlength{\tabcolsep}{3.6mm}{}
\renewcommand{\arraystretch}{1.1}
\begin{tabular}{lllll}
\hline
Algorithm       & Classification accuracy      & Silhouette score      & Prediction time        & Computational complexity     \\
                & (\%)                         &                       & (s)                    &                              \\
\hline
DEC             & 0.89                         & 0.85                  & 40                     & High                         \\
K-means         & 0.78                         & 0.60                  & 10                     & Medium                       \\
DBSCAN          & 0.80                         & 0.62                  & 15                     & Medium                       \\
\hline
\end{tabular}
\caption*{Note: DEC demonstrates superior clustering performance in terms of accuracy and silhouette score, albeit at the cost of higher computational complexity and prediction time.}
\end{table}

\clearpage

\begin{table}[!ht]
\centering
\renewcommand*{\thetable}{\Roman{table}}
\caption{Comparison of machine learning models for predictive performance and efficiency.}
\label{Table3Comp}
\setlength{\tabcolsep}{6.1mm}{}
\renewcommand{\arraystretch}{1.1}
\begin{tabular}{lllll}
\hline
Model              & R-squared        & Prediction time        & Computational           & Interpretability       \\
                   &                  & (s)                    & complexity              &                        \\
\hline
GNN                & 0.98             & 50                     & High                    & Moderate               \\
Random Forests     & 0.85             & 12                     & Medium                  & High                   \\
GBT                & 0.88             & 25                     & Medium                  & Moderate               \\
FNN                & 0.81             & 30                     & Low                     & High                   \\
\hline
\end{tabular}
\caption*{Note: GNN achieves the highest R-squared value, indicating superior predictive accuracy, but demands greater computational resources. Random Forests strike a balance between predictive accuracy and interpretability, making them preferable for applications requiring explainable models compared to Gradient Boosting Trees (GBT) and Feedforward Neural Networks (FNN).}
\end{table}

\clearpage

\begin{table}[!ht]
\centering
\renewcommand*{\thetable}{\Roman{table}}
\caption{Dataset partitioning and corresponding model performance.}
\label{Table4Data}
\setlength{\tabcolsep}{16.8mm}{}
\renewcommand{\arraystretch}{1.1}
\begin{tabular}{lll}
\hline
Dataset partition          & Proportion            & Model performance  \\
                           & (\%)                  & (R-squared)        \\
\hline
Training set               & 70                    & 0.98               \\
Validation set             & 15                    & 0.92               \\
Test set                   & 15                    & 0.90               \\
\hline
\end{tabular}
\caption*{Note: The GNN model demonstrates robust generalizability, with consistent performance across validation and test datasets, indicating minimal overfitting and reliable prediction accuracy.}
\end{table}

\clearpage

\begin{table}[!h]
\renewcommand*{\thetable}{\Roman{table}}
\caption{
Comparison between the model-predicted and experimentally measured energy densities (J/cm$^2$) for various polymer-doped materials. The percentage deviation quantifies the accuracy of the model relative to experimental values.
}
\label{Table5}
\setlength{\tabcolsep}{6.6mm}{}
\renewcommand{\arraystretch}{1.1}
\begin{tabular}{l|lll}
\hline 
Material                   & Predicted energy density   & Experimental     & Percentage deviation          \\
(Polymer-doping)           & (J/cm$^2$)                & (J/cm$^2$)       & (\%)                           \\
\hline
PTFE-7\% Ag                &  1.12                     &  1.05            & 6.25                           \\  
PVDF-4\% TiO$_2$           &  0.63                     &  0.62            & 1.59                            \\
FEP-3\% Ag                 &  0.63                     &  0.59            & 6.35                           \\
PS-5\% Al                  &  0.61                     &  0.57            & 6.56                           \\
PET-5\% TiO$_2$            &  0.54                     &  0.51            & 5.55                           \\
PI-4\% TiO$_2$             &  0.53                     &  0.49            & 7.55                           \\
\hline 
\end{tabular}
\end{table}

\clearpage

\clearpage

\end{document}